\documentclass[a4paper,12pt]{article}
\pdfoutput=1
\usepackage[pdftex]{hyperref}
\usepackage{graphicx}
\usepackage{amsmath}
\usepackage{amssymb}
\usepackage{fullpage}
\usepackage{multirow}

\def\lsim{\mathrel{\rlap{\lower4pt\hbox{\hskip1pt$\sim$}}
    \raise1pt\hbox{$<$}}}                
\def\gsim{\mathrel{\rlap{\lower4pt\hbox{\hskip1pt$\sim$}}
    \raise1pt\hbox{$>$}}}                
\newcommand{\met}{{\not\!\!E_T}} 

\begin{document}
\baselineskip 0.7cm


\begin{titlepage}
\begin{flushright}
IPMU12-0200\\
KEK-TH 1599
\end{flushright}

\vskip 1.35cm
\begin{center}

{\large\bf Top Polarization and Stop Mixing from Boosted Jet Substructure}

\vskip 1.2cm

Biplob Bhattacherjee$^1$, Sourav K. Mandal$^1$ and Mihoko Nojiri$^{1,2,3}$

\vskip 0.4cm

$^1$Kavli IPMU (WPI), University of Tokyo, Kashiwa, Chiba 277-8583, Japan\\
$^2$Theory Group, KEK, 
Tsukuba, Ibaraki 305-0801, Japan\\
$^3$The Graduate University for Advanced Studies (SOKENDAI)\\
Tsukuba, Ibaraki 305-0801, Japan\\

\vskip 1.5cm
\begin{abstract}
Top polarization is an important probe of new physics that couples to the top sector, and which may be discovered at the 14~TeV LHC.  Taking the example of the MSSM, we argue that top polarization measurements can put a constraint on the soft supersymmetry breaking parameter $A_t$.  In light of the recent discovery of a Higgs-like boson of mass $\sim 125$~GeV, a large $A_t$ is a prediction of many supersymmetric models.  To this end, we develop a {\em detector level} analysis methodology for extracting polarization information from hadronic tops using boosted jet substructure.  We show that with 100~fb$^{-1}$ of data, left and right 600~GeV stops can be distinguished to $4\sigma$, and 800~GeV stops can be distinguished to $3\sigma$.
\end{abstract}

\end{center}
\end{titlepage}

\setcounter{page}{2}


\tableofcontents

\section{Introduction}
\subsection{Motivation}
Top physics is an important probe of theories of new physics at the TeV scale, as many of these theories posit TeV-scale partners to the top quark in order to resolve the Higgs hierarchy problem.  These theories in general have chiral structure, thus a measurement of the top polarization from the decays of top partners can establish useful constraints on them.

In the case of supersymmetry with $R$-parity, the composition of the scalar top partner ``stop'' ${\tilde t}$ in terms of the weak eigenstates ${\tilde t}_R$ and ${\tilde t}_L$ can be constrained by observing the polarization of tops in the decay ${\tilde t}\rightarrow t {\tilde \chi}_1^0$.  The fermionic top partners in extra-dimensions theories with KK parity and little Higgs theories with T-parity have the analogous decays $t^{(1)}\rightarrow t B^{(1)}$  and $T'\rightarrow t B_H$.  As discrete parities are desirable because they limit dangerous contributions to electroweak precision variables~\cite{takeuchi} and admit WIMP dark matter candidates (such as ${\tilde \chi}_1^0$, $B^{(1)}$ and $B_H$), the $t+\met$ collider signature provides a useful handle on a broad class of well-motivated TeV-scale theories.

There are also numerous theories with and without top partners containing extra massive gauge bosons $Z'$ with decays such as $Z'\rightarrow t\bar t$.  Top polarization measurements can constrain the chiral structure of their couplings to the top quark.  In general, $t\bar t$ may be produced in new heavy resonances.

To date there have been a number of studies on measuring top polarization at the Large Hadron Collider (LHC).  Ref.~\cite{hisano} performed a signal-only Monte Carlo-level analysis in the context of gluino decay, and Ref.~\cite{agashe} performed a Monte Carlo-level analysis for KK gluons.  For the $t+\met$ class, Ref.~\cite{shelton} performed a signal-only parton-level formal calculation, followed by Refs.~\cite{perelstein} and \cite{berger} performing a Monte Carlo-level analysis with backgrounds, acceptance cuts and smearing effects.  For heavy resonances, Ref.~\cite{rehermann} considered the $W+$jets background and smearing effects, later Ref.~\cite{godbole} performed a signal-only Monte Carlo-level analysis, and Ref.~\cite{sakurai} studied various measurables in a fast detector simulation without backgrounds.  Ref.~\cite{krohn} elucidated the benefit of jet substructure for measuring the polarization of boosted hadronic tops for both classes of theories, although without backgrounds and at Monte Carlo-level.  

In this paper we study top polarization for the $t+\met$ class of theories at {\em detector level} including all contamination sources (e.g., ISR/FSR and MPI), relevant detector effects (e.g., magnetic field) and backgrounds.  To this end we focus on pair production of 600 GeV and 800 GeV ${\tilde t}_1$ at the 14~TeV LHC under the simplified model in which they decay entirely to $t\chi_1^0$, where $\chi_1^0\simeq {\tilde B}$ and $m_{\chi_1^0}=100$~GeV.  We choose to focus on supersymmetry also because it may be the most well-motivated of this class of theories, solving the hierarchy problem up to Planck scale as well as enhancing gauge coupling unification at high scale.

First we will briefly review the kinematics of top polarization and the phenomenology of ${\tilde t}_L$-${\tilde t}_R$ mixing.  Then we will describe our simulation and analysis methodology, and present our results for the expected sensitivity to the stop mixing angle.  Finally, we will look ahead to possibilities for improving and ramifying our methodology.

\subsection{Top polarization}  
Measurement of top polarization is possible because top quarks undergo weak decay prior to hadronization, so the top decay products carry information on the polarization of the parent quark undisturbed by the hadronization process.  The kinematics of top polarization is presented in Refs.~\cite{hisano,shelton,perelstein,godbole,kane}.  The decay products of the top quark have the angular distributions
\begin{equation}
\frac{1}{\Gamma}\frac{d\Gamma}{d(\cos\theta_{tf})}\propto 1+{\cal P}_t k_f \cos\theta_{tf}
\end{equation}
where $\cos\theta_{tf}$ is the angle between the daughter momentum and the top spin axis in the top rest frame; we can take the latter as the direction of top momentum in the lab frame.  ${\cal P}_t=\pm 1$ is the polarization of the top quark, and $k_f$ is the ``spin analyzing power'' of the daughter flavor.  For the $b$-quark,
\begin{equation}k_b = -\frac{m_t^2-m_W^2}{m_t^2+m_W^2} \simeq -0.4\end{equation}
whereas for the lepton daughter of a leptonic top decay one finds $k_l = 1$.  Consequently we can measure ${\cal P}_t$ by observing the distribution of $\cos\theta_{tf}$.  For the case of hadronic tops, this can be done directly by reconstructing the top and resolving the $b$-quark daughter.  However, for leptonic tops, one cannot fully reconstruct the top momentum due to the neutrino from the $W$ decay.  It has been proposed to define $\cos\theta_{tl}$ in an ``approximate rest frame'' in semileptonic events with reconstruction of the accompanying hadronic top~\cite{perelstein}, or require the leptonic top in a semileptonic event to be highly boosted such that one can use alternative measurables which are insensitive to top momentum in the limit $\beta
\rightarrow 1$~\cite{shelton,berger}.  Yet another proposal is to use judicious cuts to preserve polarization information in the lab frame despite the boost of the leptonic top~\cite{godbole}.

In our analysis we measure the polarization of hadronic tops, which has not only the advantage of being more simple than leptonic methods, but also that of greater statistics, as 89\% of top pairs have one hadronic top whereas only 44\% of top pairs are semileptonic~\cite{pdg}.  Moreover, leptonic top analysis entails all the difficulties of identifying isolated leptons in a real hadron collider environment.  Standard cone-based lepton isolation lose efficiency with increasing boost, hampering polarization measurements for heavy parent states.  However, this may be ameliorated by narrowing the isolation cone size as lepton $p_T$ increases~\cite{rehermann,todt}.

\subsection{Stop mixing}
As written in the review~\cite{martin}, the stop mass matrix in the weak basis under the minimal supersymmetric standard model (MSSM)  is
\begin{equation}
{\cal L}_{m_{\tilde t}}=-\left(\begin{array}{cc}\tilde t^*_L & \tilde t^*_R\end{array}\right){\bf m_{\tilde t}^2 }\left(\begin{array}{c} \tilde t_L \\ \tilde t_R\end{array}\right)
\end{equation}
\begin{equation}
{\bf m_{\tilde t}^2} = \left(\begin{array}{cc}m_{Q_3}^2 + m_t^2+\Delta_{\tilde u_L} & v(a_t^* \sin\beta - \mu y_t\cos\beta)\\v(a_t\sin\beta-\mu^*y_t\cos\beta) & m_{\bar u_3}^2+m_t^2+\Delta_{\tilde u_R}\end{array}\right)
\end{equation}
\begin{eqnarray}
\Delta_{{\tilde u}_L}&=&\left(\frac{1}{2}-\frac{2}{3}\sin^2\theta_W\right)\cos(2\beta)m_Z^2\\
\Delta_{{\tilde u}_R}&=&\frac{2}{3}\sin^2\theta_W\cos(2\beta)m_Z^2\quad .
\end{eqnarray}
If $a_t$ and $\mu$ are real, the mixing can be represented by the rotation
\begin{equation}
\left(\begin{array}{c}{\tilde t}_1 \\ {\tilde t}_2\end{array}\right)= \left(\begin{array}{cc}\cos\theta & -\sin\theta \\ \sin\theta & \cos\theta\end{array}\right)\left(\begin{array}{c}{\tilde t}_L \\ {\tilde t}_R\end{array}\right)
\end{equation}
where
\begin{equation}
\tan 2\theta = \frac{2m_t(A_t-\mu/\tan\beta)}{m_{Q_3}^2- m_{\bar u_3}^2+\Delta_{\tilde u_L}-\Delta_{\tilde u_R}}
\end{equation}
and in which $m_t=y_t v \sin\beta$ and $A_t = a_t/y_t$ have been substituted.

Let us consider the case that $\tan\beta\gg 1$ and $\mu$, $m_{Q_3}$ and $m_{\bar u_3}$ are of the same order at low scale; we also assume $m_{Q_3}>m_{\bar u_3}$ at low scale due to renormalization.
Then if $A_t = 0$ we see that ${\tilde t}_1\simeq {\tilde t}_R$ ($\theta\simeq\pi/2$); conversely, if $A_t\gsim \mu$ then ${\tilde t}_1$ will have a significant ${\tilde t}_L$ component ($\theta\sim\pi/4$).  If indeed the mixing angle $\theta$ can be measured, then under these assumptions $A_t$ may be strongly constrained.

These assumptions can be eased by incorporating other measurables.  For example, knowledge of the ${\tilde t}_1$ mass (e.g., from production rates or kinematic constraints) and the Higgs mass, which takes large corrections at one-loop that depend on $A_t$, $\mu$, $\tan\beta$, $m_{{\tilde t}_1}$ and $m_{{\tilde t}_2}$~\cite{mihoko}, would leave only the relationship between $m_{Q_3}$ and $m_{\bar u_3}$ as a model-dependent quantity. The recent discovery of a Higgs-like boson~\cite{higgsdiscovery_atlas,higgsdiscovery_cms} with a mass suggesting a large value of $A_t$ for TeV-scale supersymmetric scalars~\cite{higgsmass} may already be hinting at such a correction.  

It has also been proposed to consider the ratios of the branching fractions of stops decaying to neutralinos and charginos in order to constrain stop mixing parameters.  However, it may require input from both a hadron collider and a linear collider to have sufficient information~\cite{rolbiecki}.  In either case, this may be another set of observables which may be useful in tandem with direct polarization measurements.

To connect the stop mixing angle with top polarization, we consider the interaction term for the decay in our simplified model\footnote{For a discussion with completely general ${\tilde\chi}_1^0$, see e.g. Ref.~\cite{perelstein}.} in which $\chi_1^0\simeq \tilde B$,
\begin{eqnarray}
\notag\Delta {\cal L} &=& g'\left[\left(\frac{1}{6}\right){\tilde t}_L^*{\tilde \chi}_1^0 t_L +\left(-\frac{2}{3}\right) {\tilde t}_R^*  {\tilde \chi}_1^0t_R\right]\\
&=& g'{\tilde t}_1^*{\tilde\chi}_1^0\left[\cos\theta\left(\frac{1}{6}\right) t_L + \sin\theta \left(\frac{2}{3}\right)t_R\right]\\\notag &-&g'{\tilde t}_2^*{\tilde\chi}_1^0\left[\sin\theta\left(\frac{1}{6}\right) t_L + \cos\theta \left(\frac{2}{3}\right)t_R\right]
\end{eqnarray}
where the quantities in the parentheses are the hypercharges.  Then the observed ``effective'' mixing angle from the decay ${\tilde t}_1\rightarrow t {\tilde\chi}_1^0$ is
\begin{equation}
\tan\theta_{\rm obs} = 4\tan\theta\quad .
\end{equation}
Thus for or simplified model the stop mixing angle is amplified in the top polarization mixing angle, increasing the sensitivity to ${\tilde t}_R$ and reducing the sensitivity to small ${\tilde t}_L$ admixtures.  Under the assumptions given before, this therefore reduces the sensitivity to small values of $A_t$.

\begin{figure}[h]
\centering
\includegraphics[width=0.8\columnwidth]{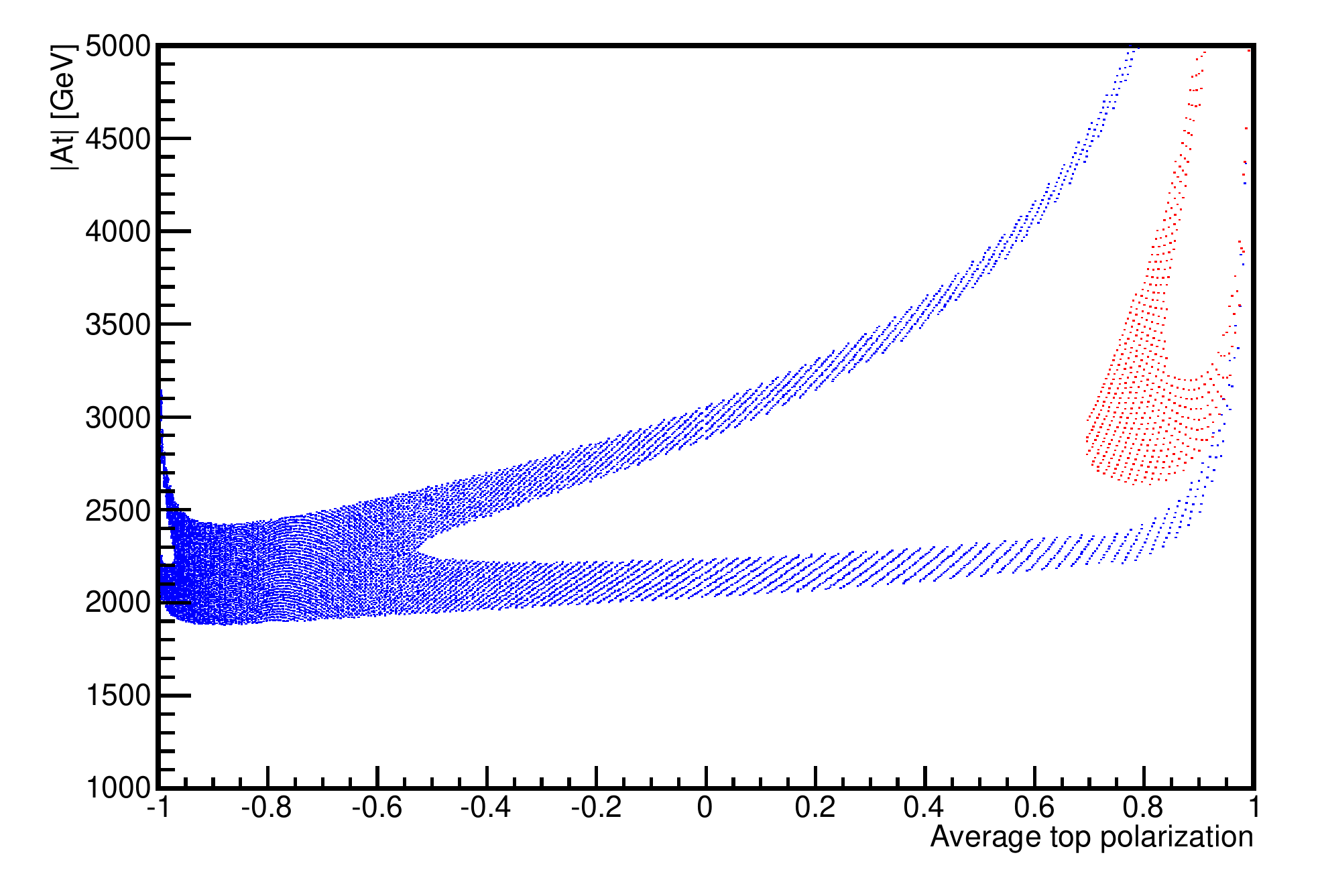}
\caption{Constraint on trilinear coupling $A_t$ vs. top polarization measurement from the MSSM Higgs mass correction for 800~GeV light stop and $125.2 < m_h < 126.2$.  Blue region is for $A_t>0$, and the red region for $A_t < 0$.}
\label{fig:susy}
\end{figure}

Putting together the Higgs mass correction and top polarization arising from stop mixing, we can see the theoretical sensitivity of $A_t$ to the measured top polarization in Figure~\ref{fig:susy}.  Here, the top polarization is defined as $(c_L^2-c_R^2)/(c_L^2+c_R^2)$, where $c_L$ and $c_R$ are the coupling strengths $(1/6)\cos\theta$ and $(2/3)\sin\theta$ to the left-handed and right-handed tops, respectively.  The Higgs mass correction is computed using {\tt FeynHiggs}~\cite{feynhiggs}.  The Higgs mass window is chosen to be the intersection of the $\pm 1\sigma$ ATLAS and CMS regions, $125.2 < m_h < 126.2$.  Indeed, sensitivity improves as the tops become more left-handed.\footnote{For this parameter scan, $m_t=173$~GeV, $\tan\beta=10$, $\mu=1692$~GeV, $m_A=1791$~GeV, $m_0=2000$~GeV, $A_b=A_\tau=1009$~GeV, $M_1=393$~GeV, $M_2=720$~GeV and $M_3=1966$~GeV, consistent with the assumptions in our discussion.}

\section{Simulation and analysis}
The goal of our analysis is to show that top polarization information can be obtained by reconstructing boosted tops in a realistic hadron collider environment.  First we describe the Monte Carlo generation of our data, then the detector simulation, and finally the reconstruction of top jets using the physics objects from the simulation.
\subsection{Event generation and detector simulation}
Using {\tt Herwig++ 2.5.0}~\cite{herwig} with all physics effects (hadronization, ISR/FSR, MPI) included, we generated left ($\sin\theta = 0$), mixed ($\tan\theta = 0.25$) and right ($\sin\theta = 1$) ${\tilde t}_1{\tilde t}^*_1$ samples with masses 600 GeV and 800 GeV for the 14~TeV LHC under our simplified model.  We calculated NLO production cross sections using {\tt Prospino2.1}~\cite{prospino}, shown in Table~\ref{table:crosssections}.

We consider the backgrounds $t\bar t$+jets, $Z$+jets, $W$+jets and $t\bar t+Z$, which we generated using {\tt MadEvent/MadGraph 5.1.3}~\cite{madgraph,matching} + {\tt PYTHIA 6.4.25}~\cite{pythia} (also with all physics effects), taking their leading order cross sections which are also shown in Table~\ref{table:crosssections}.  All extra jets are five-flavor ($g$ + $u$, $d$, $c$, $s$, $b$).
\begin{table}[h]
\centering
\begin{tabular}{|c|c|c|}
\hline
Process & Generator-level cut & Cross section \\
\hline
${\tilde t}_1 {\tilde t}_1^*$, $m=600$~GeV & --- & 218~fb\\
${\tilde t}_1 {\tilde t}_1^*$, $m=800$~GeV & --- & 36.8~fb\\
\hline
$t\bar t+\leq$ 2 jets & $p_{T,j1}>300$ GeV & 40.6 pb\\
$(Z\rightarrow \nu \bar\nu)+\leq$ 3 jets & $p_{T,j1}\,,\,\met>250$ GeV & 7.8 pb\\
$(W\rightarrow [l,\tau]\nu)+\leq$ 3 jets & $p_{T,j1}\,,\,\met>300$ GeV & 4.57 pb\\
$t\bar t+(Z\rightarrow \nu \bar\nu)$ & --- & 0.11 pb\\
\hline
\end{tabular}
\caption{Signal and background cross-sections at 14~TeV LHC.}
\label{table:crosssections}
\end{table}

For our detector simulation we used {\tt Delphes 2.0.2}~\cite{delphes}.  We modified the {\tt Delphes} codebase to use {\tt FastJet 3.0.3}~\cite{fastjet} instead of the bundled version, as the newer version has an interface to manipulate subjets at specific clustering scales or steps.  This allows us to ``prune'' the clustering tree to remove contamination, then store the resulting subjets, all within the {\tt Delphes} analysis pipeline.  

The {\tt Delphes} detector settings are tuned to ATLAS, with the hadronic calorimeter grid set to match that in the ATLAS TDR~\cite{atlastdr}.  The magnetic field is turned on in the simulation.

\subsection{Cuts}
\label{sub:cuts}
We implement the following event cuts, designed to increase the significance $S/\sqrt{S+B}$ for our characteristic $t+\met$ signature:
\begin{enumerate}
\item {\em $\met >$300~GeV.}
\item {\em Leading fat jet $p_T >$400~GeV.}
\item {\em If there are no leptons w/ $p_T>5$~GeV, require subleading fat jet $p_T>$100~GeV.}  This cut suppresses processes like $W/Z$+jets in which the second fat jet is from QCD, as these jets are likely to be soft.  Since the majority of signal events have leptons due to $W\rightarrow l\nu$ and the decay of $b$-flavored mesons, we require that there are no leptons for this cut.  Thus, this cut is most effective against $Z$+jets.
\item {\em Lepton is not collimated with $\met$}.  For every lepton with $p_T>5$~GeV, require
\begin{equation}
\frac{\cos(\phi_\met - \phi_l)}{(\met+p_{T,l})/(350\;{\rm GeV})} < 0.4\quad .
\end{equation}
This selects against high $\met$ arising from boosted leptonic $W$ decays in $t\bar t$+jets and $W$+jets, as the opening angle between the lepton and $\vec\met$ is likely to be much smaller in these processes than from a top partner decay.
\item {\em Hard subjet is not collimated with $\met$}.  For every subjet with $p_T>50$~GeV, require the same as above.  This works against hadronic $\tau$ from $W$ decays in $t\bar t$+jets and $W$+jets, as well as highly collimated $b$-subjets from top decays in $t{\bar t}+$jets.
\item {\em Require at least one top tagged jet}, using the procedure described in the next section.
\item {\em $225< {M_T}_2 < 650$ for 600~GeV stops, and $325 < {M_T}_2 < 850$ for 800~GeV stops.}  ${M_T}_2$ is calculated for the leading and subleading jet, with $m_\chi=0$.  For the leading jet we used the reconstructed top jet if top tagged, otherwise we used the trimmed jet; similarly for the subleading jet.  We employed the ${M_T}_2$ code of Ref.~\cite{ucdavismt2}.  
\end{enumerate}
The resulting cut flow for signal events is shown in Table~\ref{table:cutflowsignal}, and for background events in Table~\ref{table:cutflowbackground} for 14~TeV LHC @ 100~fb$^{-1}$.  The cuts reveal some preference for right stops, which is noted in Ref.~\cite{perelstein}.
\begin{table}[h]
\centering
\begin{tabular}{|c|c|c|c|c|c|c|c|}
\hline
\multicolumn{2}{|c|}{\multirow{2}{*}{Cut}} & \multicolumn{3}{|c|}{Stop 600 GeV} & \multicolumn{3}{|c|}{Stop 800 GeV}\\
\cline{3-8}
\multicolumn{2}{|c|}{} & Left & Mixed & Right & Left & Mixed & Right\\
\hline
\# & Pre-cut & 21900 & 21900 & 21900 & 3680 & 3680 & 3680\\
\hline\hline
1 & $\met > 300$~GeV & 9513 & 9739 & 9857 & 2394 & 2411 & 2433\\
2 & $p_{T,j1} > 400$~GeV & 5415 & 5496 & 5472 & 	1816 & 1835 & 1825\\
3 & If $n_l=0$, $p_{T,j2}>100$~GeV & 5150 & 5220 & 	5192 & 1756 & 1773 & 1764\\
4 & lepton/$\met$ collimation & 4155 & 4315 & 4394 &	1515 & 1555 & 1559\\
5 & subjet/$\met$ collimation & 2914 & 3046 & 3084 & 1171 & 1209 & 1208\\
\hline
6 & \# top tag $\geq 1$ & 1014 & 1065 & 1082 & 450 & 456 & 463\\
7a & $225< {M_T}_2 < 650$ & 908 & 954 & 969 & --- & 	--- & ---\\
7b & $325 < {M_T}_2 < 850$ & --- & --- & --- & 364 & 	369 & 374\\
\hline
\end{tabular}
\caption{Cut flow for signal events at 14~TeV LHC @ 100 fb$^{-1}$.}
\label{table:cutflowsignal}
\end{table}

\begin{table}[h]
\centering
\begin{tabular}{|c|c|c|c|c|c|c|}
\hline
\multicolumn{2}{|c|}{Cut}  & $t{\bar t}+$jets & $Z+$jets & $W+$jets & $t{\bar t}+Z$ \\
\hline
\# & Generator-level & $4.06\times 10^6$ & $7.8\times 10^5$ & $4.57\times 10^5$ & 11000\\
\hline\hline
1 & $\met > 300$~GeV & $1.30\times 10^5$ & $4.96\times 10^5$ & $4.36\times 10^5$ & 815\\
2 & $p_{T,j1} > 400$~GeV & 90503 & $2.28\times 10^5$ & $3.20\times 10^5$ & 351\\
3 & If $n_l=0$, $p_{T,j2}>100$~GeV & 88133 & $1.01\times 10^5$ & $2.68\times 10^5$ & 326\\
4 & lepton/$\met$ collimation & 21518 & 98441 & 69865 & 237\\
5 & subjet/$\met$ collimation & 4412 & 60860 & 37852 &	149 \\
\hline
6 & \# top tag $\geq 1$ & 1140 & 305 & 99 & 52\\
7a & $225< M_{T2} < 650$ & 554 & 222 & 56 & 43\\
7b & $325 < M_{T2} < 850$ & 275 & 141 & 45 & 30\\
\hline
\end{tabular}
\caption{Cut flow for background events at 14~TeV LHC @ 100 fb$^{-1}$.}
\label{table:cutflowbackground}
\end{table}

\subsection{Top jet reconstruction}
Many aspects of this analysis are well-reviewed in Ref.~\cite{tilmantop}.

\subsubsection{Jet clustering and grooming}
Hadrons are clustered as fat jets using the Cambridge/Aachen algorithm~\cite{cajets} with cone size $R=1.2$ and subjet cone size $\Delta R = 0.2$.  These numbers are chosen such that the distribution in the number of subjets per fat jet peaks at $\sim 3$ in our signal samples after the following grooming procedures:
\begin{enumerate}
\item The jet clustering trees are ``pruned''~\cite{pruning} using the mass-drop condition~\cite{massdrop}
\begin{equation}
m_{j_{n-1}} < 0.8 \times m_{j_n}
\end{equation}
where $m_{j_{n-1}}$ is the invariant mass of the hardest parent jet, and $m_{j_n}$ is the invariant mass of the child jet at clustering step $n$.  We also require the subjet separation condition
\begin{equation}
d_{k_T}(j_{n-1,1}, j_{n-1,2}) > (\Delta R)^2\cdot m_{j_n}^2
\end{equation}
where $d_{k_T}$ is the $k_T$ distance between the two parents jets.  This removes contamination from MPI and ISR, improving top reconstruction quality.  We take the fat jet (and its subjets) at the clustering step where both conditions are satisfied.
\item This jet is then ``trimmed''~\cite{trimming}, removing subjets with $p_T<10$~GeV.  This further reduces contamination. 
\end{enumerate}

The trimmed jet is then fed to our top reconstruction algorithm.
\subsubsection{Reconstruction and tagging}
The following algorithm is attempted for every fat jet:
\begin{enumerate}
\item Require $p_T>400$~GeV for the {\em untrimmed} jet.
\item Require at least three subjets.
\item Require that one $b$-subjet $j_b$ is reconstructed in the jet.
\item Find the subjet combination $j_1 j_2 j_b$, of which no pair of subjets are within $\Delta R$ of each other, and which gives the closest invariant mass to $m_t$.  Require also that this invariant mass be in the window (150, 200)~GeV.
\item If there is no successful tag, require at least four subjets and retry the step above with four-subjet combinations $j_1 j_2 j_3 j_b$.
\item Optionally require one two-subjet combination to have an invariant mass in the loose $W$ mass window (50, 110)~GeV.  We present our main results both with and without this requirement.
\end{enumerate}
This algorithm differs from the Johns Hopkins top tagger~\cite{hopkinstagger} by declustering more than two steps in the pruning stage if necessary, using a mass drop condition rather than a $p_T$ drop condition, having an absolute rather than fractional trimming threshold,  requiring a $b$-tag instead of imposing a $W$ mass condition, and not imposing a top helicity angle condition (as this is our observable).  The algorithm also differs from the CMS tagger~\cite{cmstagger} by not requiring a minimum two-subjet invariant mass.

Our algorithm differs also from the HEPTopTagger~\cite{heptoptagger} by not imposing the various two-subjet mass requirements, having a mandatory $b$-tag, and also by implementing four-subjet reconstruction.

We make these choices to enhance top tagging efficiency, presuming that the top parent particle has already been discovered.  Nonetheless, top mistagging does not overwhelm the signal as will be apparent in our results.

\subsubsection{$b$-tagging subjets}
\label{sec:btagging}
We require $b$-tagging in our top reconstruction algorithm to reconstruct the observable $\cos\theta_{tb}$ with high fidelity, but also to suppress top mistagging from background processes.  Utilizing recent advances in $b$-tagging by the LHC detector collaborations, we employ the $b$-tagging efficiencies (shown in Table~\ref{table:btag}) recently validated at 7~TeV LHC by CMS for their CSVM tagger~\cite{cmsbtagging}.  We impose the upper limit $p_T = 1000$~GeV to be conservative, though it is not indicated by the CMS study.  We choose to use the CMS efficiencies since they are validated up to $p_T=670$~GeV, though a recent ATLAS study~\cite{atlasbtagging} shows similar efficiencies up to 200~GeV.

\begin{table}
\centering
\begin{tabular}{|c|c|}
\hline
Kinematic region & Efficiency\\
\hline
$p_T < 30$ GeV & 0\%\\
$30\;{\rm GeV} < p_T < 60\;{\rm GeV}$ & 60\%\\
$60\;{\rm GeV} < p_T < 450\;{\rm GeV}$ & 70\%\\
$450\;{\rm GeV} < p_T < 1000\;{\rm GeV}$ & 60\%\\
$1000\;{\rm GeV}<p_T$ & 0\%\\
\hline
\end{tabular}
\caption{Utilized $b$-tagging efficiencies.}
\label{table:btag}
\end{table}

However, we {\em do not} implement mistagging, as this depends on various factors that are best implemented by experimenters --- mistagging rates vary rapidly with tagging efficiency and so are sensitive to systematic uncertainties that we cannot model.  We expect charm jets to contribute to most of the mistags.  For example, in Ref.~\cite{cmsbtagging} a 60\% $b$-tagging efficiency {\em nominally} results in a 10\% charm mistagging rate in Monte Carlo, whereas for 50\% $b$-tagging efficiency this drops to 4\%.  The mistagging rate for lighter flavors is $\sim 1$\%, so this contribution is negligible.

We apply these $b$-tagging efficiencies at parton level.  To reconstruct a $b$-subjet we sum all the subjets within $\Delta R = 0.2$ of the $b$-parton, as this captures some hard FSR.  If more than one $b$-parton yields a matching subjet inside a given fat jet, that fat jet is rejected and is not top tagged.

\section{Results}
\subsection{Reconstruction quality}
Our measurable is $\cos\theta_{tb}$, where $\theta_{tb}$ is the angle between the $b$ daughter and the top spin axis in the top rest frame.  The usefulness of this measurable depends on the fidelity of the top jet reconstruction.  Here we illustrate the effectiveness of our reconstruction method.

\begin{figure}[h]
\includegraphics[width=\columnwidth]{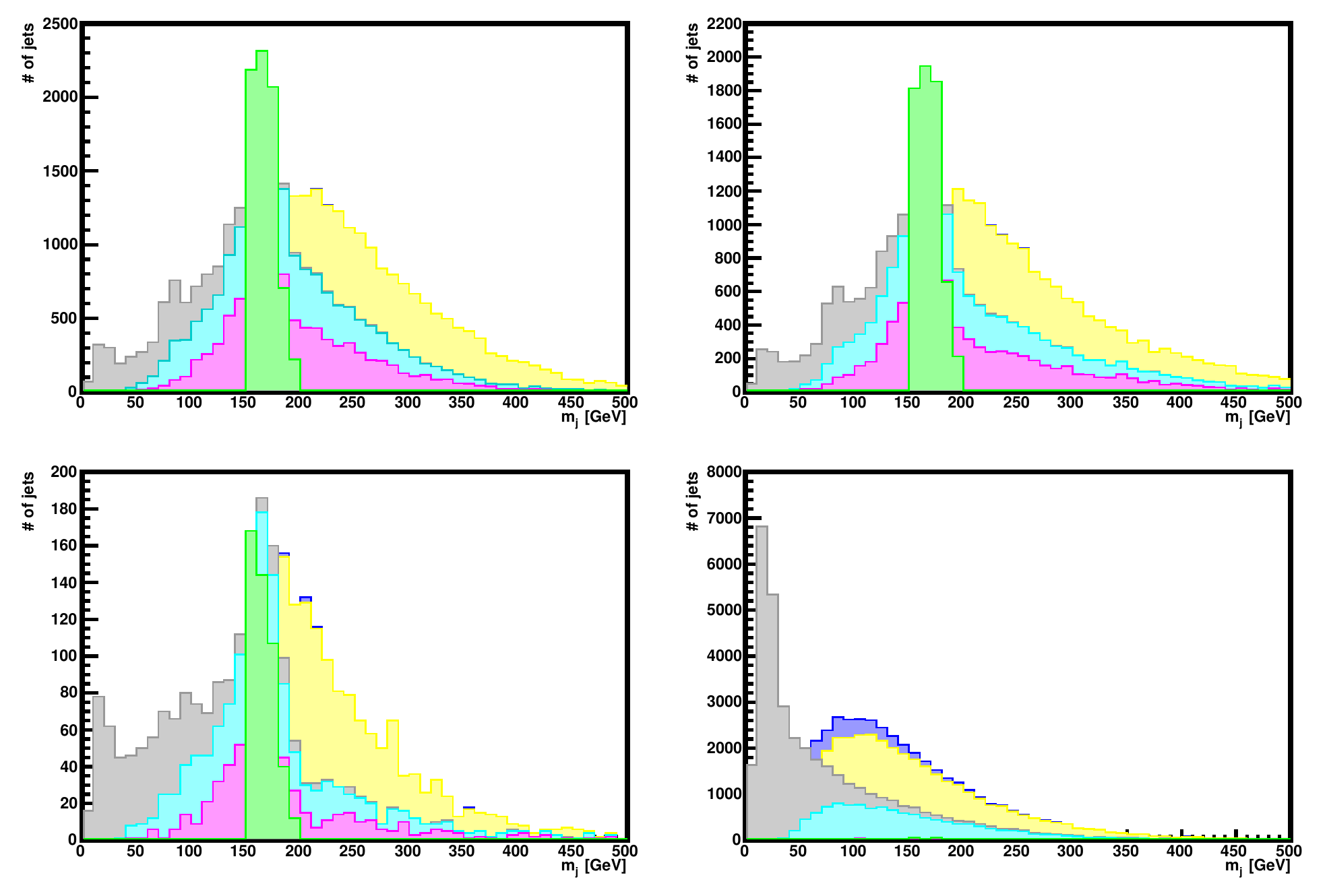}
\caption{Jet invariant mass sequence for 600~GeV mixed stops (upper left), 800~GeV mixed stops (upper right), $t\bar t$+jets (lower left) and $Z$+jets (lower right).  In each panel, from back to front, shown is the jet mass distribution after $p_T>400$~GeV cut (blue), requiring successful pruning (yellow), trimming (gray), requiring three or more subjets (cyan), $b$-tagging (magenta), and finally top tagging (green).}
\label{fig:jet_invm}
\end{figure}

First, we show the jet invariant mass distribution at different stages in Figure~\ref{fig:jet_invm}.  For signal processes and $t\bar t$+jets, few jets are lost in the pruning phase; however, without jet cluster pruning, we find that the $\cos\theta_{tb}$ distribution loses fidelity compared to the parton-level expectation.  The subsequent trimming step is essential for removing soft contamination, resulting in a shift of the invariant mass lower towards the correct top mass.  Then, requiring at least three subjets and a $b$-tag narrows the distribution further.  Finally, top reconstruction assembles the subjets with correct invariant mass.  Modulo $b$-tagging efficiency, the top tagging efficiency for hadronic tops in our samples which pass the $p_T$ cut is $\sim60$\%.  Conversely, top reconstruction suppresses $Z$+jets and $W$+jets by a factor of $O(100)$.

\begin{figure}[h]
\includegraphics[width=\columnwidth]{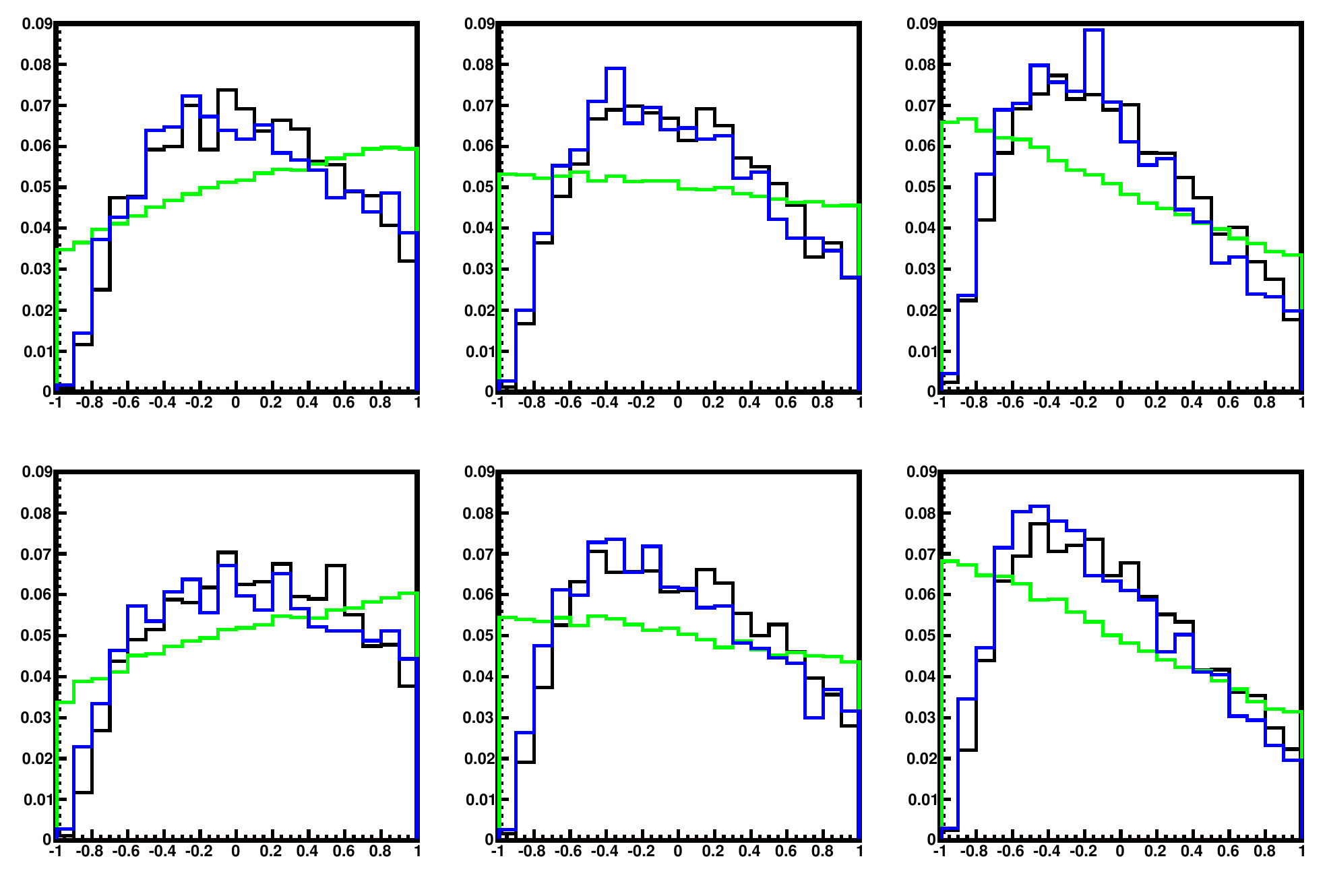}
\caption{ $\cos\theta_{tb}$ distributions for stops only, normalized to one.  Left to right are left, mixed, and right stop samples; upper row is for 600~GeV stops, the lower for 800~GeV stops.  Shown in each panel are the reconstructed distribution (black), parton-level distribution (blue), and parton-level distribution before cuts and tagging (green).}
\label{fig:ctb_stops}
\end{figure}

The quality of reconstruction is also apparent in the signal-only $\cos\theta_{tb}$ distributions shown in Figure~\ref{fig:ctb_stops}.  One sees that the parton-level and reconstructed distributions coincide up to statistical fluctuations.   Near $\cos\theta_{tb}=-1$ there is depletion due to poor $b$-tagging efficiency, whereas near $\cos\theta_{tb}=+1$ there is depletion due to the $W$ subjets being too soft to pass the trimming threshold.  One sees for this reason that there is less contrast between left and mixed stops with mass 600~GeV than with mass 800~GeV. The distributions at pre-cut/tag parton-level match Ref.~\cite{shelton}, except they have a downward left-to-right tilt due to the harder $b$-partons losing energy to FSR.

\begin{figure}[h]
\includegraphics[width=\columnwidth]{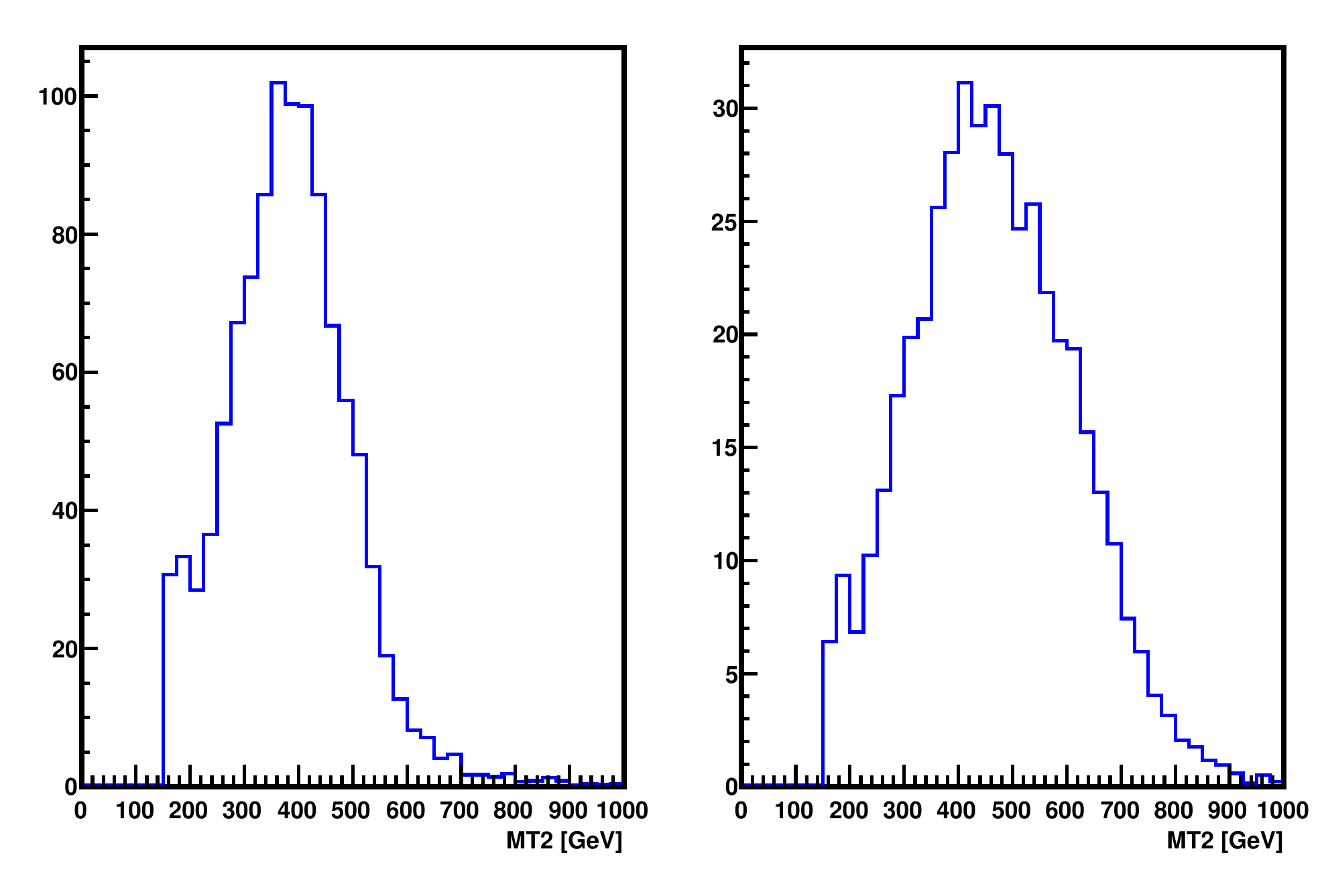}
\caption{Sample ${M_T}_2$ distributions for mixed stops at reconstruction level, signal only.  Left panel is for stop mass 600~GeV, and the right panel is for mass 800~GeV.}
\label{fig:mt2}
\end{figure}

Finally, we show sample ${M_T}_2$ distributions for 600~GeV and 800~GeV mixed stops in Figure~\ref{fig:mt2} at reconstruction level, signal only.  One can clearly see the expected inflection points at  ${M_T}_2=600$~GeV and ${M_T}_2=800$~GeV.  As described in Subsection~\ref{sub:cuts}, we consider only top-tagged events, calculating ${M_T}_2$ with the two leading fat jets; for each we use the reconstructed jet if it is top-tagged, otherwise we use the trimmed jet.

\subsection{$t\bar t$ + jets control region}
\begin{figure}[h]
\includegraphics[width=\columnwidth]{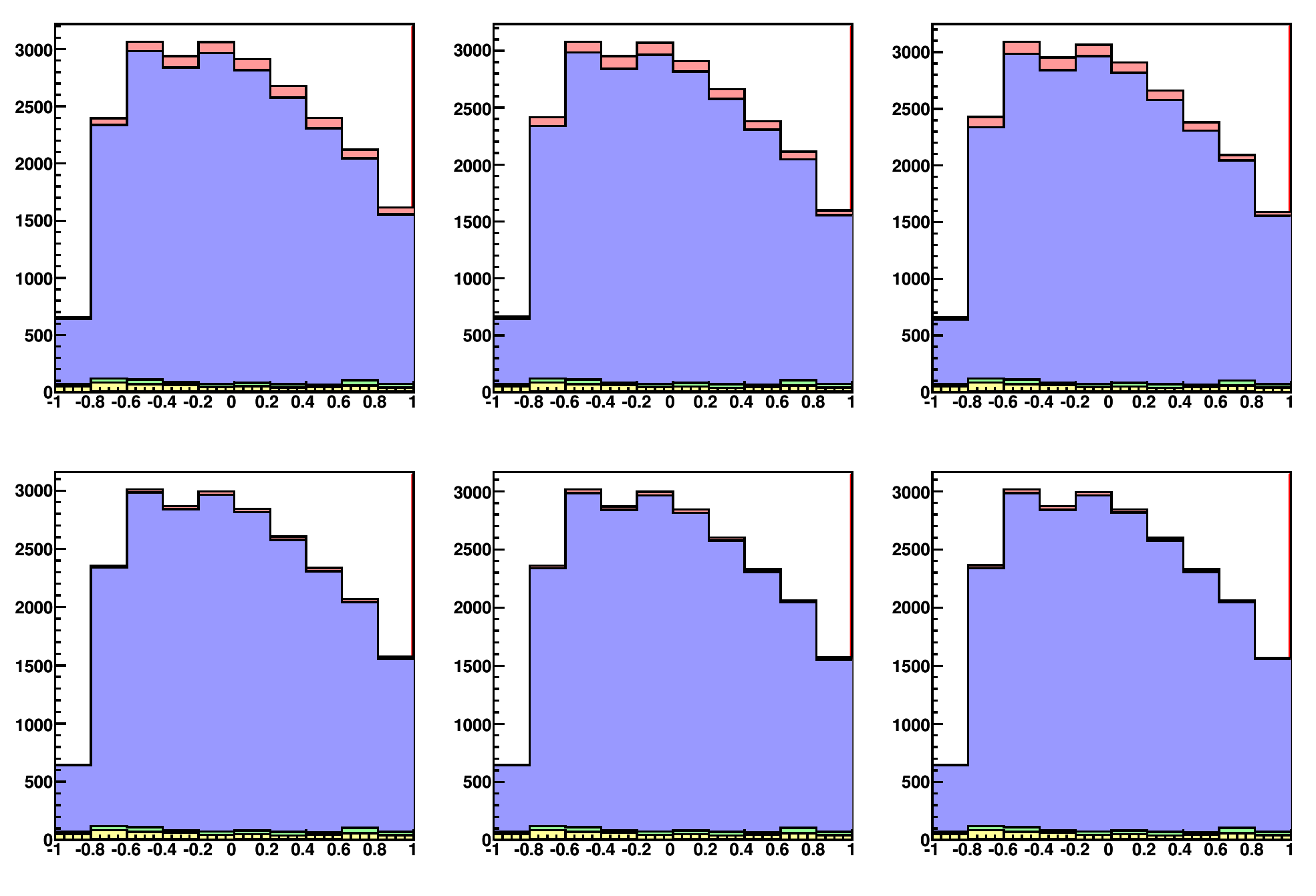}
\caption{Event distributions of $\cos\theta_{tb}$ for $t\bar t+$jets in the control region at 14~TeV LHC @ 100~fb$^{-1}$.  Left-to-right are left, mixed and right stop samples; upper row is for 600~GeV stops, the lower row for 800~GeV stops.  Shown in each panel top to bottom is signal (red), ${t\bar t}+$jets (blue), $Z+$jets (green), $W$+jets (yellow), and $t\bar t+Z$ (gray; hardly visible).}
\label{fig:ttcontrol}
\end{figure}

As experimenters are likely to rely on data rather than Monte Carlo for the $t\bar t$+jets background, we demonstrate a control region by inverting cuts \#4 and \#5, {\em requiring} that there be at least one lepton or hard subjet highly collimated with $\vec\met$. An array of $\cos\theta_{tb}$ distributions for this region is shown in Figure~\ref{fig:ttcontrol}.  They evince little contamination from signal and other backgrounds.

\begin{figure}[h]
\centering
\includegraphics[width=0.8\columnwidth]{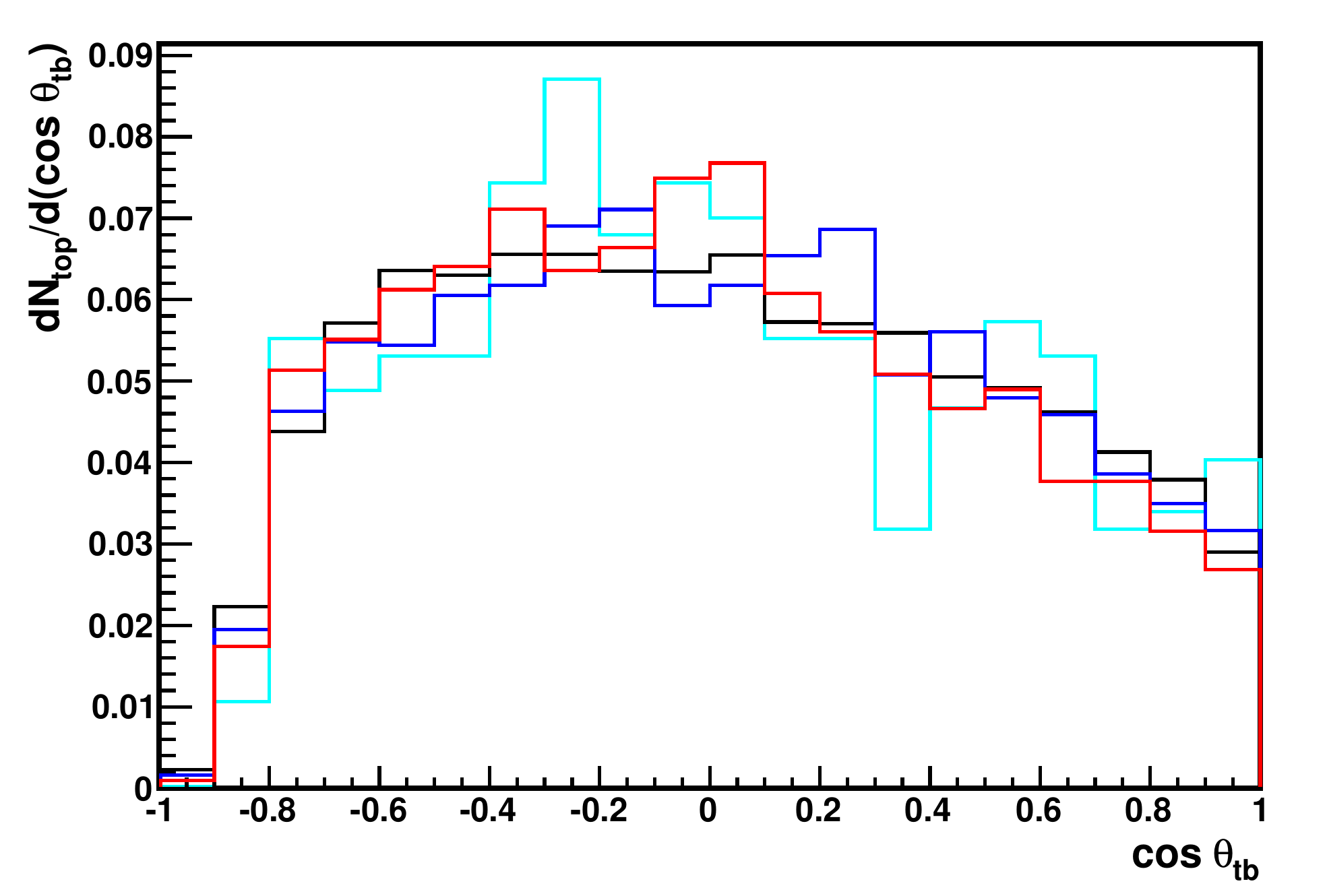}
\caption{Distributions of $\cos\theta_{tb}$ for ${t\bar t}$+jets, normalized to one.  Shown are the distribution without collimation cuts (black), with the lepton collimation cut only (blue), with the subjet collimation cut only (red), and with both collimation cuts applied (cyan).}
\label{fig:ttcutcheck}
\end{figure}

\begin{figure}[h]
\centering
\includegraphics[width=0.8\columnwidth]{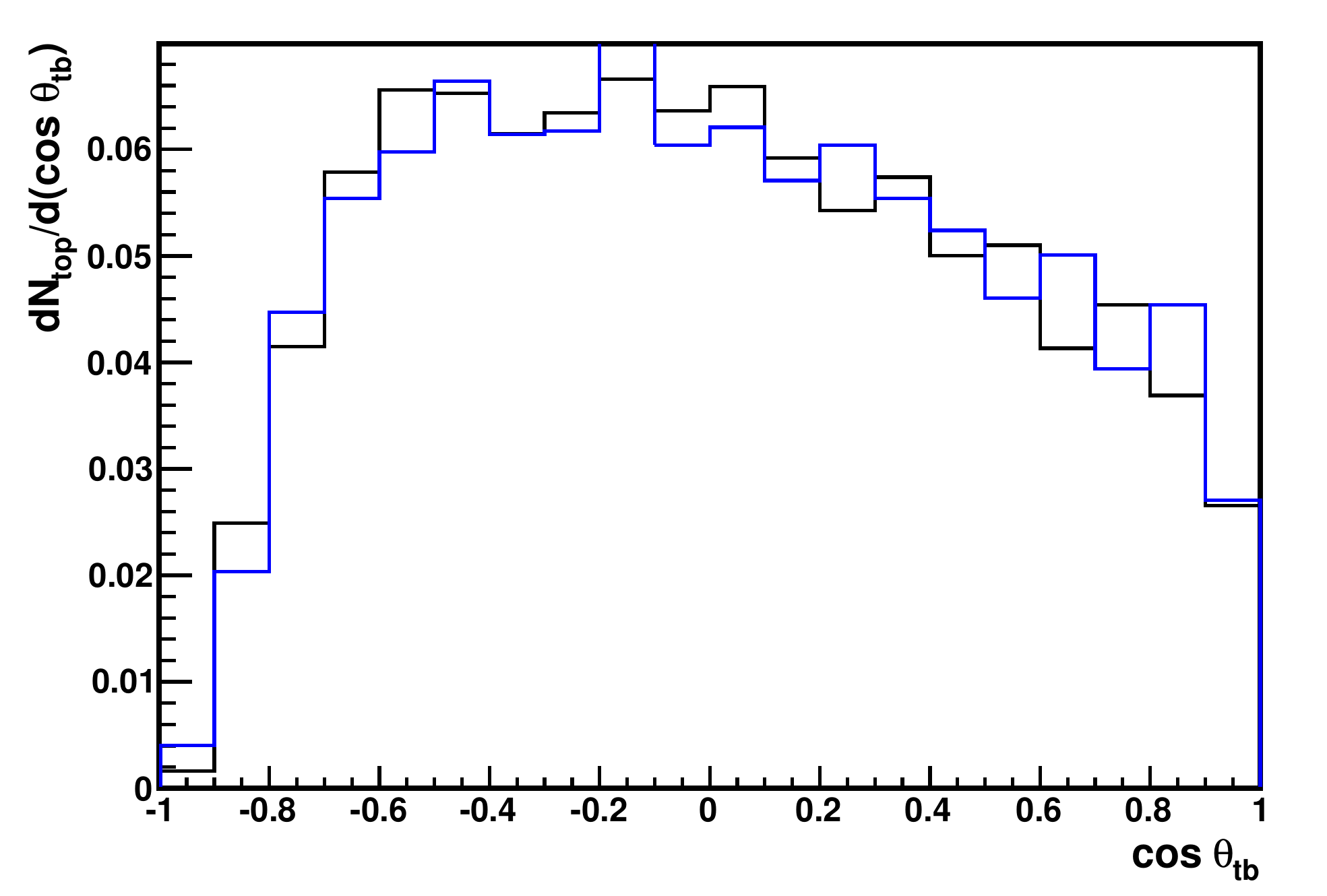}
\caption{Distributions of $\cos\theta_{tb}$ for the ${t\bar t}$+jets control sample, normalized to one.  Shown are the distribution for ${M_T}_2< 225$~GeV (black), and for ${M_T}_2 > 225$~GeV.}
\label{fig:tthilo}
\end{figure}

It may be of concern that the collimation cuts \#4 and \#5 might introduce distortions to the $\cos\theta_{tb}$ distributions obtained from top tagging.  In Figure~\ref{fig:ttcutcheck} we show that the polarization distribution for $t\bar t$+jets is invariant under these cuts up to statistical fluctuations.  We show also in Figure~\ref{fig:tthilo} that the ${M_T}_2$ cut does not distort the polarization distribution of $t\bar t$+jets in the control region.  Thus, it is safe to use a data-driven $t\bar t$+jets background from this region.

\subsection{Sensitivity to stop mixing}
\begin{figure}
\includegraphics[width=\columnwidth]{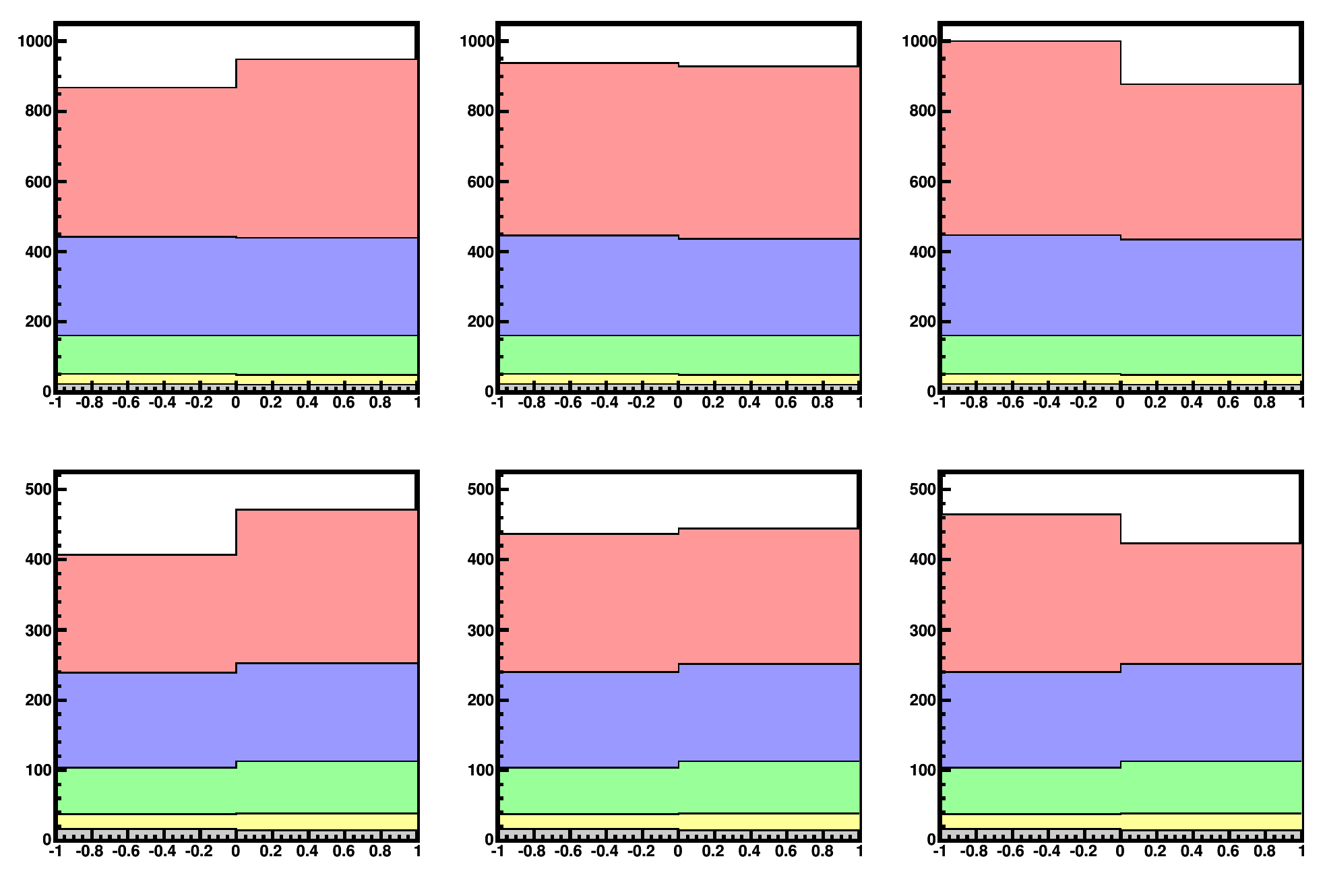}
\caption{Sum of $\cos\theta_{tb}$ distributions from different processes for 14~TeV LHC @ 100~fb$^{-1}$, {\em without} a $W$ mass condition. Left to right are left, mixed and right stop samples; upper row is for 600~GeV stops, and the lower is for 800~GeV stops.  Shown are the contributions from signal (red), $t\bar t$+jets (blue), $Z+$jets (green), $W+$jets (yellow), $t\bar t + Z$ (gray).}
\label{fig:combostack}
\end{figure}

To calculate sensitivity, we sum the $\cos\theta_{tb}$ distributions from signal and background processes.  For the $t\bar t+$jets contribution, we use the control region distribution normalized to the total of the signal region distribution.  This sum is shown in Figure~\ref{fig:combostack}, where we have rebinned the data to use only two bins in order to minimize the trials penalty.  The corresponding $p$-values for distinguishing stop mixing hypotheses are shown in Table~\ref{table:pvalues}.  For 600~GeV stops at 14~TeV LHC @ 100~fb$^{-1}$, left and right mixtures can be distinguished to better than $4\sigma$, and left/right can be distinguished from the mixed state to better than $1.5\sigma$.  For 800~GeV stops, left and right can be distinguished to nearly $3\sigma$.

\begin{table}[h]
\centering
\begin{tabular}{|l||c|c|c||c|c|c|}
\hline
\multirow{2}{*}{{\em Truth}} &  \multicolumn{3}{|c||}{{\em  Hypothesis ($m_{{\tilde t}_1}=$600 GeV)}} &  \multicolumn{3}{|c|}{{\em  Hypothesis ($m_{{\tilde t}_1}=$800 GeV)}}\\
\cline{2-7}
 & left & mixed & right & left & mixed & right\\
\hline
left & 1 & 0.047 & $2.6\times10^{-6}$ & 1 & 0.16 & 0.0014\\
mixed & 0.058 & 1 & 0.030 & 0.17 & 1 & 0.24\\
right & $8.1\times 10^{-6}$ & 0.032 & 1 & 0.0018 & 0.25 & 1\\
\hline
\end{tabular}
\caption{$p$-values for distinguishing stop mixing hypotheses at 14~TeV LHC @ 100~fb$^{-1}$, with no $W$ mass condition.}
\label{table:pvalues}
\end{table}

\begin{figure}
\includegraphics[width=\columnwidth]{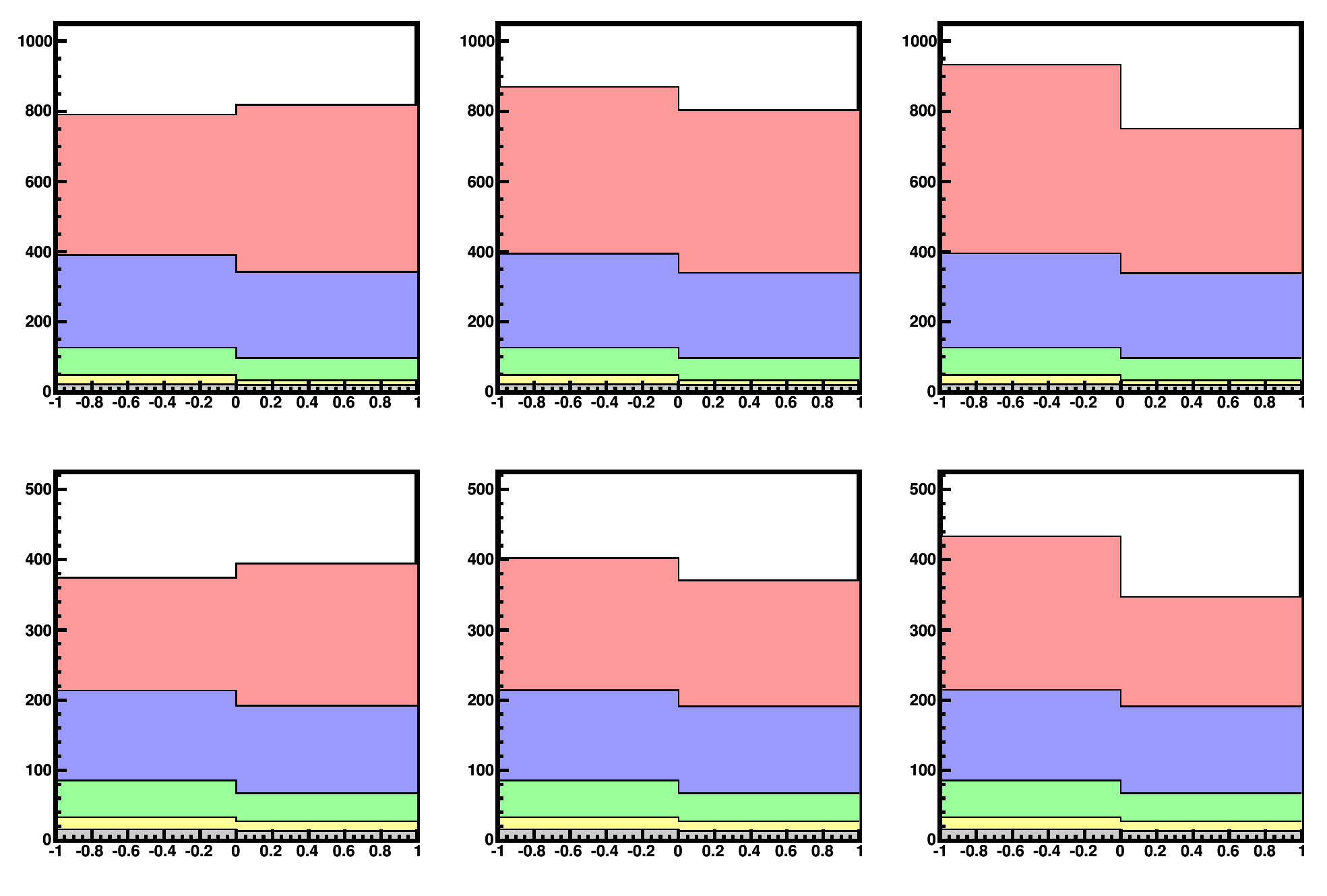}
\caption{Sum of $\cos\theta_{tb}$ distributions from different processes for 14~TeV LHC @ 100~fb$^{-1}$ {\em with} a $W$ mass condition. Left to right are left, mixed and right stop samples; upper row is for 600~GeV stops, and the lower is for 800~GeV stops.  Shown are contributions from signal (red), $t\bar t$+jets (blue), $Z+$jets (green), $W+$jets (yellow), $t\bar t + Z$ (gray).}
\label{fig:combostack_mW}
\end{figure}

Except for $t\bar t +Z$, most of the tops from the background processes surviving the cuts and reconstruction are due to extra QCD jets carrying $b$-flavor.  The results therefore are improved somewhat by imposing a loose $W$ mass condition in the top reconstruction algorithm.  If we require the mass of some two-subjet combination (not including the $b$-subjet) to be in the window $(50, 110)$~GeV, we obtain the sum plot shown in Figure~\ref{fig:combostack_mW} and the corresponding $p$-values in Table~\ref{table:pvalues_mW}.  For 600~GeV stops, left and right can now be distinguished to better than $4.5\sigma$, and left/right can be distinguished from mixed to roughly $2\sigma$.  For 800~GeV stops, left and right can now be distinguished to better than $3\sigma$.  However, imposing this $W$ mass condition may increase systematic errors from the $W/Z+$jets backgrounds due to the greater variation between the negative and positive bins.  This occurs because the subjets of the fake $W$s have lower $p_T$ at larger $\cos\theta_{tb}$, and are less likely to satisfy the $W$ mass condition.

We conclude that stop mixing hypotheses can be distinguished at the 14~TeV LHC with $\sim100$~fb$^{-1}$ of data.

\begin{table}[h]
\centering
\begin{tabular}{|l||c|c|c||c|c|c|}
\hline
\multirow{2}{*}{{\em Truth}} &  \multicolumn{3}{|c||}{{\em  Hypothesis ($m_{{\tilde t}_1}=$600 GeV)}} &  \multicolumn{3}{|c|}{{\em  Hypothesis ($m_{{\tilde t}_1}=$800 GeV)}}\\
\cline{2-7}
 & left & mixed & right & left & mixed & right\\
\hline
left & 1 & 0.018 & $1.7\times10^{-6}$ & 1 & 0.17 & $5.6\times 10^{-4}$\\
mixed & 0.025 & 1 & 0.017 & 0.17 & 1 & 0.14\\
right & $9.0\times 10^{-7}$ & 0.018 & 1 & $7\times 10^{-4}$ & 0.15 & 1\\
\hline
\end{tabular}
\caption{$p$-values for distinguishing stop mixing hypotheses at 14~TeV LHC @ 100~fb$^{-1}$, with a $W$ mass condition.}
\label{table:pvalues_mW}
\end{table}

\section{Conclusion and outlook}
In summary, after reviewing the motivation for top polarization measurement, the kinematics of top polarization, and the phenomenology of stop mixing, we described a simulation and analysis methodology for the $t+\met$ collider signature that can distinguish stop mixing hypotheses at 14~TeV LHC with $\sim 100$~fb$^{-1}$ of data.

There are several possible improvements to the methodology.  The first and foremost is to include polarization information from leptonic decays, perhaps using the techniques of~\cite{rehermann,todt}.  Also if, in consultation with experimenters, the trimming threshold of $p_T =10$~GeV or the subjet cone size $\Delta R = 0.2$ can be reduced, this would enhance the performance of the top reconstruction algorithm.  Finally, it may be useful to implement the top reconstruction technique of Ref.~\cite{krohn} to supplement $b$-tagging, especially at large boosts.

Other possible improvements are to:
\begin{itemize}
\item Implement charm mistagging in the $b$-tagging algorithm, though this may be non-trivial as discussed in Section~\ref{sec:btagging}.
\item Use spin-correlated backgrounds, e.g. from {\tt ALPGEN}~\cite{alpgen}.  We did produce a small sample of $t\bar t+$jets in {\tt ALPGEN} but found no difference in the $\cos\theta_{tb}$ distribution from the {\tt MadGraph/MadEvent}+{\tt PYTHIA} events used in this analysis.
\item Calculate backgrounds at NLO using, e.g., {\tt MC@NLO}~\cite{mcatnlo} to increase accuracy.  However, this is likely beyond our computational capabilities unless it is used only to normalize the leading-order backgrounds.  We estimate that the $K$-factor for the $t\bar t$ background to be less than 1.2 given our cuts $p_{T,j1}>400$ and $\met > 300$~GeV~\cite{melnikov}, though there is evidence that with an additional jet one may actually have $K < 1$~\cite{dittmaier}.
\end{itemize}

In conclusion, the LHC running at 14~TeV may provide new discoveries in the top sector, in which case top polarization will be an important tool for constraining this new physics.  The methodology in this paper may be useful for this purpose.

\section*{Acknowledgments}
This work is supported by the World Premier International Research Center Initiative (WPI Initiative) as well as Grant-in-Aid for Scientific Research Nos. 22540300, 23104005 from the Ministry of Education, Science, Sports, and Culture (MEXT), Japan.  The authors would like to thank T.~T.~Yanagida and S.~Matsumoto for preliminary discussions on this topic.  We would also like to thank Z.~Heng, K.~Sakurai and C.~Wymant for useful comments.



\begin{thebibliography}{99}



\bibitem{takeuchi}
T.~Takeuchi, O.~Lebedev and W.~Loinaz,
``Constraints on R-parity violation from precision electroweak measurements,''
[hep-ph/0009180].

\bibitem{hisano}
J.~Hisano, K.~Kawagoe and M.~M.~Nojiri,
Phys.\ Rev.\ D {\bf 68}, 035007 (2003)
[hep-ph/0304214].

\bibitem{agashe}
K.~Agashe, A.~Belyaev, T.~Krupovnickas, G.~Perez and J.~Virzi,
 Phys.\ Rev.\ D {\bf 77}, 015003 (2008)
  [hep-ph/0612015].

\bibitem{shelton}
J.~Shelton,
Phys.\ Rev.\ D {\bf 79}, 014032 (2009)
[arXiv:0811.0569 [hep-ph]].

\bibitem{perelstein}
M.~Perelstein and A.~Weiler,
JHEP {\bf 0903}, 141 (2009)
[arXiv:0811.1024 [hep-ph]].

\bibitem{berger}
 E.~L.~Berger, Q.~-H.~Cao, J.~-H.~Yu and H.~Zhang,
Phys.\ Rev.\ Lett.\  {\bf 109}, 152004 (2012)
[arXiv:1207.1101 [hep-ph]].

\bibitem{rehermann}
K.~Rehermann and B.~Tweedie,
JHEP {\bf 1103}, 059 (2011)
[arXiv:1007.2221 [hep-ph]].

\bibitem{godbole}
 R.~M.~Godbole, K.~Rao, S.~D.~Rindani and R.~K.~Singh,
JHEP {\bf 1011}, 144 (2010)
[arXiv:1010.1458 [hep-ph]].

\bibitem{sakurai}
A.~Papaefstathiou and K.~Sakurai,
JHEP {\bf 1206}, 069 (2012)
[arXiv:1112.3956 [hep-ph]].

\bibitem{krohn}
D.~Krohn, J.~Shelton and L.~-T.~Wang,
JHEP {\bf 1007}, 041 (2010)
[arXiv:0909.3855 [hep-ph]].

\bibitem{kane}
 G.~L.~Kane, G.~A.~Ladinsky and C.~P.~Yuan,
Phys.\ Rev.\ D {\bf 45}, 124 (1992).

\bibitem{pdg}
J.~Beringer et~al. (Particle Data Group), Phys. Rev. D86, 010001 (2012).

\bibitem{todt}
S.~Todt,
``Top quark resonances in ATLAS simulation,'' (DESY Summer Student Program, 2012).


\bibitem{martin}
S.~P.~Martin,
``A Supersymmetry primer,''
In ``Kane, G.L. (ed.): Perspectives on supersymmetry II'' 1-153
[hep-ph/9709356].

\bibitem{mihoko}
M.~Drees and M.~M.~Nojiri,
Phys.\ Rev.\ D {\bf 45}, 2482 (1992).

\bibitem{higgsdiscovery_atlas}
G.~Aad {\it et al.}  [ATLAS Collaboration],
Phys.\ Lett.\ B {\bf 716}, 1 (2012)
[arXiv:1207.7214 [hep-ex]].

\bibitem{higgsdiscovery_cms}
S.~Chatrchyan {\it et al.} [CMS Collaboration],
Phys.\ Lett.\ B {\bf 716}, 30 (2012)
[arXiv:1207.7235 [hep-ex]].

\bibitem{higgsmass}
H.~Baer, V.~Barger and A.~Mustafayev,
Phys.\ Rev.\ D {\bf 85}, 075010 (2012)
[arXiv:1112.3017 [hep-ph]];
 S.~Heinemeyer, O.~Stal and G.~Weiglein,
Phys.\ Lett.\ B {\bf 710}, 201 (2012)
[arXiv:1112.3026 [hep-ph]].A.~Arbey, M.~Battaglia, A.~Djouadi, F.~Mahmoudi and J.~Quevillon,
Phys.\ Lett.\ B {\bf 708}, 162 (2012)
[arXiv:1112.3028 [hep-ph]];
 O.~Buchmueller, R.~Cavanaugh, A.~De Roeck, M.~J.~Dolan, J.~R.~Ellis, H.~Flacher, S.~Heinemeyer and G.~Isidori {\it et al.},
Eur.\ Phys.\ J.\ C {\bf 72}, 2020 (2012)
[arXiv:1112.3564 [hep-ph]];
  J.~-J.~Cao, Z.~-X.~Heng, J.~M.~Yang, Y.~-M.~Zhang and J.~-Y.~Zhu,
JHEP {\bf 1203}, 086 (2012)
[arXiv:1202.5821 [hep-ph]];
 C.~Wymant,
Phys.\ Rev.\ D {\bf 86}, 115023 (2012)
[arXiv:1208.1737 [hep-ph]].

\bibitem{rolbiecki}
K.~Rolbiecki, J.~Tattersall and G.~Moortgat-Pick,
Eur.\ Phys.\ J.\ C {\bf 71}, 1517 (2011)
[arXiv:0909.3196 [hep-ph]].

\bibitem{feynhiggs}
S.~Heinemeyer, W.~Hollik and G.~Weiglein,
Eur.\ Phys.\ J.\ C {\bf 9}, 343 (1999)
[hep-ph/9812472];
S.~Heinemeyer, W.~Hollik and G.~Weiglein,
Comput.\ Phys.\ Commun.\  {\bf 124}, 76 (2000)
[hep-ph/9812320];
 G.~Degrassi, S.~Heinemeyer, W.~Hollik, P.~Slavich and G.~Weiglein,
Eur.\ Phys.\ J.\ C {\bf 28}, 133 (2003)
[hep-ph/0212020];
 M.~Frank, T.~Hahn, S.~Heinemeyer, W.~Hollik, H.~Rzehak and G.~Weiglein,
JHEP {\bf 0702}, 047 (2007)
[hep-ph/0611326].

\bibitem{herwig}
M.~Bahr {\it et al.},
  ``Herwig++ Physics and Manual,''
  Eur.\ Phys.\ J.\  C {\bf 58} (2008) 639
  [arXiv:0803.0883 [hep-ph]].

\bibitem{prospino}
W.~Beenakker, M.~Kramer, T.~Plehn, M.~Spira and P.~M.~Zerwas,
Nucl.\ Phys.\ B {\bf 515}, 3 (1998)
[hep-ph/9710451].

\bibitem{madgraph}
  J.~Alwall, M.~Herquet, F.~Maltoni, O.~Mattelaer and T.~Stelzer,
  JHEP {\bf 1106}, 128 (2011)
  [arXiv:1106.0522 [hep-ph]].

\bibitem{matching}
  M.~L.~Mangano, M.~Moretti, F.~Piccinini and M.~Treccani,
production in hadronic collisions,''
  JHEP {\bf 0701}, 013 (2007)
  [hep-ph/0611129].

\bibitem{pythia}
  T.~Sjostrand, S.~Mrenna and P.~Z.~Skands,
  JHEP {\bf 0605}, 026 (2006)
  [hep-ph/0603175].

\bibitem{delphes}
S.~Ovyn, X.~Rouby and V.~Lemaitre,
[arXiv:0903.2225 [hep-ph]].

\bibitem{fastjet}
  M.~Cacciari, G.~P.~Salam and G.~Soyez,
Eur.\ Phys.\ J.\ C {\bf 72}, 1896 (2012)
[arXiv:1111.6097 [hep-ph]].

\bibitem{atlastdr}
``ATLAS: Detector and physics performance technical design report. Volume 1,'' CERN-LHCC-99-1.

\bibitem{tilmantop}
T.~Plehn and M.~Spannowsky,
J.\ Phys.\ G {\bf 39}, 083001 (2012)
[arXiv:1112.4441 [hep-ph]].

\bibitem{cajets}
 Y.~L.~Dokshitzer, G.~D.~Leder, S.~Moretti and B.~R.~Webber,
JHEP {\bf 9708}, 001 (1997)
[hep-ph/9707323].

\bibitem{pruning}
S.~D.~Ellis, C.~K.~Vermilion and J.~R.~Walsh,
Phys.\ Rev.\ D {\bf 80}, 051501 (2009)
[arXiv:0903.5081 [hep-ph]];  S.~D.~Ellis, C.~K.~Vermilion and J.~R.~Walsh,
Phys.\ Rev.\ D {\bf 81}, 094023 (2010)
[arXiv:0912.0033 [hep-ph]].

\bibitem{massdrop}
 J.~M.~Butterworth, B.~E.~Cox and J.~R.~Forshaw,
Phys.\ Rev.\ D {\bf 65}, 096014 (2002)
[hep-ph/0201098].

\bibitem{trimming}
 D.~Krohn, J.~Thaler and L.~-T.~Wang,
JHEP {\bf 1002}, 084 (2010)
[arXiv:0912.1342 [hep-ph]].

\bibitem{hopkinstagger}
D.~E.~Kaplan, K.~Rehermann, M.~D.~Schwartz and B.~Tweedie,
Phys.\ Rev.\ Lett.\  {\bf 101}, 142001 (2008)
[arXiv:0806.0848 [hep-ph]].

\bibitem{cmstagger}
 [CMS Collaboration],
 ``A Cambridge-Aachen (C-A) based Jet Algorithm for boosted top-jet tagging,''
CMS-PAS-JME-09-001.

\bibitem{heptoptagger}
 T.~Plehn, M.~Spannowsky, M.~Takeuchi and D.~Zerwas,
JHEP {\bf 1010}, 078 (2010)
[arXiv:1006.2833 [hep-ph]].

\bibitem{cmsbtagging}
[CMS Collaboration],
``b-Jet Identification in the CMS Experiment,''
CMS-PAS-BTV-11-004.

\bibitem{atlasbtagging}
[ATLAS Collaboration],``Measurement of the $b$-tag Efficiency in a Sample of Jets Containing Muons with 5~fb$^{-1}$ of Data from the ATLAS Detector,'' ATLAS-CONF-2012-043.

\bibitem{ucdavismt2}
H.~-C.~Cheng, J.~F.~Gunion, Z.~Han, G.~Marandella and B.~McElrath,
JHEP {\bf 0712}, 076 (2007)
[arXiv:0707.0030 [hep-ph]];
  H.~-C.~Cheng and Z.~Han,
JHEP {\bf 0812}, 063 (2008),
[arXiv:0810.5178 [hep-ph]].

\bibitem{alpgen}
 M.~L.~Mangano, M.~Moretti, F.~Piccinini, R.~Pittau and A.~D.~Polosa,
JHEP {\bf 0307}, 001 (2003)
[hep-ph/0206293].

\bibitem{mcatnlo}
S.~Frixione and B.R.~Webber, 
 JHEP 0206 (2002) 029 [hep-ph/0204244].

\bibitem{melnikov}
 K.~Melnikov and M.~Schulze,
JHEP {\bf 0908}, 049 (2009)
[arXiv:0907.3090 [hep-ph]].

\bibitem{dittmaier}
S.~Dittmaier, P.~Uwer and S.~Weinzierl, Eur.\ Phys.\ J.\ C 
59, 625 (2009) [arXiv:0810.0452 [hep-ph]].

\end{thebibliography}




\end{document}